\documentstyle[psfig]{l-aa}

\def\sectionApp#1{ \refstepcounter{section}\section*{Appendix \Alph{section}: 
#1}}
\def\LabelFig#1#2{ \refstepcounter{figure}\count100=\thefigure		
	           \def\thefigure{\the\count100 #1}\label{#2}	
                   \addtocounter{figure}{-1}}				

\begin{document}

\thesaurus{09(06.09.1;06.15.1;06.18.2;03.13.4)}
%
\title{Inferring the equatorial solar tachocline from frequency splittings}
%
\author{T. Corbard \and G. Berthomieu \and J. Provost \and  P. Morel}
\institute{Laboratoire G.-D. Cassini, CNRS UMR 6529, Observatoire de la C\^ote 
d'Azur, BP 4229, 06304 
 Nice Cedex 4, FRANCE }
\offprints{T. Corbard}
\date{Received 18 July 1997; accepted 24 October 1997}
\maketitle

\begin{abstract}
%
Helioseismic inversions, carried out for several years on various
ground-based and spatial observations, have shown that the solar
rotation rate presents two principal regimes: a quasi-rigid rotation
in the radiative interior and a latitude-dependent rotation in the whole
convection zone. The thin layer, named solar tachocline, between these
two regimes is difficult to infer through inverse techniques because 
of the ill-posed nature of the problem that requires 
regularization techniques
which, in their global form,
 tend to smooth out any high gradient in the solution. Thus, most of the 
previous attempts to study the rotation profile of the solar tachocline
have been carried out through forward modeling. In this work we show
that some appropriate inverse techniques can also be used and we 
  compare the ability of three 1D inverse techniques combined with
 two automatic strategies for the choice of the regularization parameter, 
to infer the solar tachocline profile in the equatorial plane.
 Our work,
 applied on LOWL (LOWL is an abbreviation for low degree denoted by L)
two years dataset, 
argue in favor of a very sharp ($0.05\pm0.03R_\odot$) transition zone 
located at $0.695\pm0.005R_\odot$ which is in good agreement with the previous
forward analysis carried out on Global Oscillations Network Group (GONG),
Big Bear Solar Observatory  (BBSO) and LOWL datasets.

\end{abstract}

\section{Introduction}
%
Helioseismic inversions of the solar p-modes frequencies splitted
by rotation have shown that there is, at the base of the convection zone, a
thin transition layer separating two regimes of rotation, a strong differential
rotation in the convection zone and a quasi rigid rotation in the radiative
 interior (e.g. Thompson et al. \cite{Science}; Corbard et al. \cite{meAA}).
 This layer, called tachocline, is supposed to play an important role in the solar dynamo, in the transport of angular momentum and in the mixing of chemical elements.
Its position  $r_c$ and thickness $w$ give constraints to the theories 
describing its structure and evolution (Spiegel \& Zahn \cite{tacho1};
Gough \& Sekii \cite{tacho2}). Different estimations of these parameters
have been obtained so far mostly by using forward methods
(Kosovichev \cite{koso}; Charbonneau et al. \cite{char}; Basu \cite{Basu}).

The aim of this work is to test and compare the ability of some
inversion methods to infer the location and the width of 
the solar tachocline, and then to apply these methods to
 helioseismic data. 
We compare three 1D least-squares methods. They differ essentially by the
mean  used to regularize the ill-posed inverse problem of inferring the
equatorial solar rotation rate from the observed frequency splittings. 
The first method is the most commonly used Regularized Least-Squares (RLS)
 method
with Tikhonov regularization (\cite{Tikhonov}),
 the second one is the Modified Truncated Singular Value Decomposition (MTSVD)
introduced by Sekii and Shibahashi (\cite{MTSVD1})  
which uses a regularization term of the
same form but with a discrete truncation parameter instead of  the continuous
Tikhonov regularization parameter. The third method, introduced by Hansen \&
Mosegaard ({\cite{PPTSVD}), 
is called
Piecewise Polynomials TSVD (PP-TSVD) and is a modification of the 
MTSVD method that can preserve discontinuities of the solution. 

In Sect.~\ref{sec:Model}, we briefly recall the inverse problem and define 
our parameterization of the tachocline. Section~\ref{sec:methods} gives
the two strategies studied in this work for inferring the rapid 
variation of the rotation.
We test these methods by inverting artificial data in Sect.~\ref{sec:tests} 
and then, in Sect.~\ref{sec:lowl}, we use 
this study in order to infer the location and thickness of the solar
 tachocline in the equatorial plane  from data observed
 by the LOWL instrument 
(Tomczyk et al. \cite{Tomczyk}).


\section{Direct analysis and parameterization of the tachocline}
\label{sec:Model}
%
Frequency splittings $\Delta\nu_{nlm}=\nu_{nlm}-\nu_{nl-m}$ between 
modes with the same radial order $n$ and degree $l$ but different azimuthal
orders $m$ are induced by the solar rotation
$\Omega(r,\theta)$ expressed as a function of the radius $r$ and colatitude 
$\theta$. For a slow rotation, assumed to be  symmetric about the equator, 
and moderate or high degree modes,
 these splittings are given by: 
 
\begin{equation}\label{eq:int2D}
\Delta\nu_{nlm}\!=m\!\int_0^{{\pi\over 2}}\!\!\!\!\int_0^{R_\odot}\!\!\!
 K_{nl}(r)P_l^m(\cos\theta)^2 \ \Omega(r,\theta)\ 
\sin\theta\ dr\ d\theta,
\end{equation}
where $K_{nl}(r)$ are the so-called rotational kernels that can be
calculated for each mode from a  solar model (Morel et al. \cite{updated}).
 In the following, 
they are assumed to be known exactly. There exists additional terms that are 
not taken into account in Eq.~(\ref{eq:int2D}) but, 
as discussed in Corbard et al. (\cite{meAA}), they 
do not influence inversion above
$0.4R_\odot$. As the aim of this work is not to sound the
rotation of the core, Eq.  (\ref{eq:int2D}) is a good approximation.
$P_l^m(\cos\theta)$ are normalized
Legendre functions. Their asymptotic property leads, as discussed by
Antia et al. (\cite{Antia}), to the following expression that
shows  the sectoral (i.e. $l=m$) modes splittings as weighted
averages of the  equatorial rotation rate 
$\Omega_{eq}(r)=\Omega(r,90^{\circ})$:

\begin{equation}\label{eq:int}
\Delta\nu_{nll}\simeq l\int_0^{R_\odot} K_{nl}(r)\ \Omega_{eq}(r)\ dr.
\end{equation} 
We note that
 the validity of this 1D approximation is $l$-dependent. Indeed, the higher
the degree, the more 
the latitudinal kernel $P_l^l(\cos\theta)^2\sin\theta$ is  peaked
at the equator.

Following Charbonneau et al. (\cite{char}), we
 define the location and the width of the transition 
zone in the equatorial plane 
as the parameters $\hat r_c$ and $\hat w$ respectively of the following $erf$
 function which fits the rotation law in this plane:

\begin{equation}\label{eq:erf}
\Omega_{eq}(r)=\hat\Omega_0+{1\over 2}(\hat\Omega_1-\hat\Omega_0)
\left(1+erf\left({r-\hat r_c\over0.5 \hat w}\right)\right).
\end{equation} 
Here $\hat \Omega_0$ and $\hat \Omega_1$ represent the mean values 
of the rotation in the
radiative interior and in the convection zone respectively.

In order to compare different 1D inverse methods, we have built 
several sets of theoretical sectoral frequency splittings that correspond
to different given  rotation laws 
with fixed  parameters 
$r_c$, $w$, $\Omega_0$,  $\Omega_1$   but with 
a function of 
the colatitude in order to mimic the latitudinal differential rotation
 of the convection zone:

\begin{equation}\label{eqn:law}
\Omega(r,\theta)\!\!=\!\!\Omega_0\!+\!{1\over 2}
(\Omega_1\!-\!A\!\cos^2\!\!\theta\!-\!B\!\cos^4\!\!\theta-\Omega_0)\!\!
\left(\!\!1\!\!+erf\!\!\left(\!\!{r-r_c\over 0.5w}\!\!\right)\!\right)
\end{equation} 
Evidently, for any choice of constants $A$ and $B$, the searched parameters
for these rotation laws are $\hat r_c=r_c$, $\hat{w} =w$, 
$\hat\Omega_0=\Omega_0$ and $\hat\Omega_1=\Omega_1$.
 We compute the  splittings $\Delta\nu_{nll}$  from Eq.~(\ref{eq:int2D})
 for a set of modes corresponding to the
set of LOWL data used in Corbard et al. (1997)
 and
we add a normally distributed noise 
$\delta\nu_{nll}\in {\cal N}(0,\sigma_{nl})$.
For each mode $(n,l)$ the standard deviation of
the noise $\sigma_{nl}$ has been taken equal to:

\begin{equation}\label{eq:err}
\sigma_{nl}={\bar\sigma_{nl}\over\sqrt{k_\sigma}}, 
\end{equation}
where $\bar\sigma_{nl}$ is the error derived from the observers' 
uncertainties  for
a splitting $\Delta\nu_{nll}$,
and $k_\sigma$ is an integer used to vary the level of the noise that we 
introduce in the data. Doing this, we take into account the fact that
the error obtained on the observed splitting varies with the frequency 
and the degree of the
mode which  is certainly more realistic
than taking the same average standard deviation for all the modes.
From those noisy splittings, the equatorial rotation profile is obtained by 
inverting Eq.~(\ref{eq:int}) and this profile  
is then fitted by the
$erf$ function Eq.~(\ref{eq:erf}) leading to the parameters 
$\bar r_c$, $\bar w$, $\bar\Omega_0$, $\bar\Omega_1$ 
which will be compared to the initial parameters.

\section{Strategies for inferring rapid variations of the rotation}
\label{sec:methods}

The three inverse methods used in this work are detailed in Appendix 
\ref{app:methods}. They all use a grid of $50$ points 
in radius distributed according to the density of turning points of observed 
modes. 
The most important difficulty in inferring the thickness of the tachocline 
from inverse methods results from the fact that the problem of solving
Eq.~(\ref{eq:int}) is an ill-posed problem and this is strengthened by the
fact that rotational kernels give redundant 
information about the outer 
part of the sun whereas they have only low amplitude
in the solar core for the observed mode set.
 Numerically, this produces  a
 high value for the condition number
(defined as the maximum singular value divided by the smallest singular value)
of the discretized problem Eq.~(\ref{eq:discret}) 
(typically $\Lambda_{max}/\Lambda_{min} 
\simeq 2\times 10^8$ in our implementation) 
and the singular values decay rapidly.
This high value of the condition number means that the solution of the 
initial problem is highly
sensitive to the numerical errors and the noise contained in the data.
 Therefore we have to introduce
some a-priori knowledge on the rotation profile. Unfortunately this
regularization tends to smooth out every rapid 
variation in the solution. By using global regularization, 
we make the implicit assumption
that the real rotation is smooth everywhere and therefore the information
 about the thickness of a rapid variation of the rotation profile is not 
directly readable from the solutions obtained by classic inversions.
There are however several ways for overcoming these difficulties.

\subsection{Local deconvolution of the result obtained from linear inversions:
the use of averaging kernels}
The first way 
is to have a good understanding of the process by which the inversion
smoothes the solution: using this information, we may  be  able to
inverse  this process and to acquire a more realistic view of the rotation.
This is what Charbonneau et al. (\cite{char}) have done 
in combination  with
the so-called Subtractive Optimal Localized Average (SOLA)
(Pijpers \& Thompson \cite{SOLA1}, \cite{SOLA2})  method.
This can be generalized
for any linear inversion as RLS method used in this work.
 The solution $\bar\Omega(r_0)$ obtained at a target location $r_0$ can 
be viewed as a weighted average of the `true rotation' $\Omega(r)$,
 the weighting function being the averaging kernel $\kappa(r,r_0)$ 
that can always be estimated at any $r_0$:

\begin{equation}\label{eq:avk}
\bar\Omega(r_0)=\int_0^{R_\odot}\kappa(r,r_0)\Omega(r)\ dr. 
\end{equation}
If we suppose that 
the averaging kernels 
obtained at any depth can be  approximated by a translation of
the averaging kernel obtained at the middle of the transition i.e. 
$\kappa(r,\hat r_c)$,
then we can define $\kappa_c$ by $\kappa_c(r-\hat r_c)\equiv\kappa(r,\hat r_c)$ and
Eq.~(\ref{eq:avk}) reduces to a convolution equation:

\begin{equation}
\bar\Omega(r_0)=\!\!\int_0^{R_\odot}\!\!\kappa_c(r-r_0)\Omega(r)\ dr \Leftrightarrow 
\bar\Omega(r)=\kappa_c(r)*\Omega(r)
\end{equation}
Finally, if the `true rotation'  
can be  well approximated by an $erf$ function of the
form given by Eq.~(\ref{eq:erf}), and
if we approximate the kernel $ \kappa_c(r-r_0)$ by a Gaussian
function of the form:

\begin{equation}
 \kappa_c(r-r_0)\simeq \exp\left[-(r-r_0)^2/\Delta_r^2\right],
\end{equation}
then the inferred solution is also an $erf$ function of the form 
Eq.~(\ref{eq:erf}) but with a larger width $\bar{w}$. 
A simple deconvolution gives the following relation between 
the searched width $\hat{w}$ and the inferred width $\bar{w}$:

\begin{equation}\label{eq:correction}
\hat{w}=\bar{w}_c\equiv\sqrt{{\bar{w}}^2-4\Delta_r^2},
\end{equation}
which defined the corrected inferred width $\bar{w}_c$.

This result is valid only under a large number of assumptions that
may be quite distant from the reality.
Especially the reduction to a convolution form is certainly not
valid because of the extent of averaging kernels that tend to increase rapidly
toward the solar core. Moreover the profile of the rotation rate
 may be much more complicated than a simple $erf$ function.
However, the tachocline  is thin and the averaging kernels have 
nearly the same profile in its whole extent. 
Thus this is certainly a good approach to get a
quantitative idea of how the inversion enlarges the `true rotation' transition.
We note that if we obtain $\Delta_r>\bar{w}/2$ this certainly means that
some of the previous assumptions are not valid.
In this work, we have applied this `deconvolution method' on the solutions
obtained by Tikhonov inversions computed as explained in Appendix \ref{app:Ti}.
We estimate that this cannot be made for MTSVD method because the 
corresponding averaging kernels are less well peaked and exhibit 
a more oscillatory behavior (see Fig. \ref{fig:avk} hereafter). 

\subsection{Non linear regularization}
The second way to estimate the location and thickness of the tachocline, is to build inverse methods that are capable
of producing solutions with steep gradients. The idea is
to apply a local regularization instead of the global 
Tikhonov regularization term. This leads to a non linear
problem and piecewise smooth solutions. This approach has
recently found useful applications in image processing
for edge-preserving regularization (Aubert et al. \cite{i3s}) and
total variation (TV) denoising (Vogel \& Oman \cite{Vogel1}, \cite{Vogel2}).
 In particular,
 the TV of $f$ is defined as the 1-norm of the first derivative
of $f$ and this is the definition of smoothness that we use
in the PP-TSVD inverse method. Therefore, the results obtained by this method,
detailed in Appendix \ref{app:M-PPTSVD}, 
represent a first attempt to use this
class of inversion with non linear regularization  on helioseismic data.

\section{Tests with artificial data: results and discussion}\label{sec:tests}
\subsection{The key: how to choose regularization parameters}\label{sec:key}
Whichever regularized inverse method we use, a very important point
is the choice of the regularization parameter which can be
a discrete truncation parameter $k$ (MTSVD, PP-TSVD, Eq.~(\ref{eq:tsvd})) or a 
continuous
parameter $\lambda$ (Tikhonov, Eq.~(\ref{eq:Ti})). This choice is specially important
if we want to infer a quantity like the width of a zone with high
gradients which  is directly affected by the regularization.
Several methods for choosing the regularization
parameter have been proposed that tend to establish a balance between
the propagation of input errors and the regularization 
(see e.g. Badeva \& Morozov (\cite{Morozov}), Thompson \& Craig (\cite{ThompsonAM2})
 and Hansen (\cite{L-curve}, \cite{HansenTools}) 
for a general review and Thompson (\cite{ThompsonAM1}), Barett (\cite{Barrett}) and 
Stepanov \& Christensen-Dalsgaard (\cite{Stepanov}) for applications in helioseismic inversions).
 In this work we test and compare the ability of two
of these automatic strategies,
 namely the L-curve criterion (Hansen \cite{L-curve}) 
and the Generalized Cross 
Validation (GCV) criterion (Wahba \cite{Wahba}; Golub et al. \cite{Golub}),
 to reproduce a good estimation of the tachocline
profile from noisy data.

The importance of the choice of the regularization 
parameter can be illustrated by the following figures 
(Figs. \ref{fig:good}, \ref{fig:bad}, \ref{fig:goodk}, \ref{fig:badk} )
where the
results of the fit of the solution by an $erf$ function are plotted
as a function of the regularization parameter.  

\begin{figure} \LabelFig{a-d}{fig:good}
\psscalefirst
\centerline{
\psfig{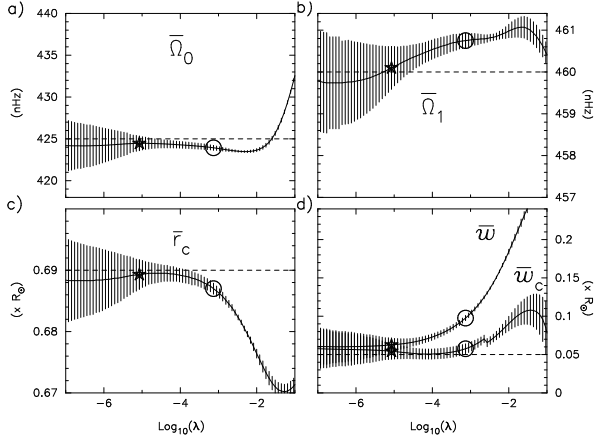}}
 \caption[]{Inferred parameters $\bar\Omega_0$, $\bar\Omega_1$, $\bar r_c$, 
$\bar{w}$ and corrected inferred parameter $\bar w_c$ against the 
logarithm of the Tikhonov regularization parameter $\lambda$. Error
bars result from the fit of the solution by an $erf$ function taking into 
account the propagation of noise through the inverse process but not the
existing correlations between the results obtained  at two different radius.
 The initial
parameters are indicated by dashed lines. The GCV and L-curve choices
are shown by the full star and the circle respectively. The input rotation law 
was not dependent on the latitude ($A=B=0$) and the level of noise was small 
($k_\sigma=10$).}
\end{figure}

Figure \ref{fig:good} represents the variation of  the
 four $erf$-parameters $\bar\Omega_0$, $\bar\Omega_1$,
$\bar r_c$ and $\bar\omega$ deduced from a Tikhonov inversion
as a function of the logarithm of the regularization  parameter. 
 The four initial
parameters were $\Omega_0=425$ nHz, $\Omega_1=460$ nHz, $r_c=0.69R_\odot$ and
$w=0.05R_\odot$. In this case, called the `ideal case' in the following, 
the added errors were small ($k_\sigma$=10)
and the initial rotation law was not dependent on the latitude ($A=B=0$).
 The choices designated by L-curve and GCV strategies
are shown by the full star and the circle respectively.
 In addition we have plotted the 
corrected inferred width $\bar{w}_c$
given by Eq.~(\ref{eq:correction}) and computed by
calculating systematically the averaging kernel at $r_0=\bar r_c$ 
(as shown on Fig. \ref{fig:avk}a for the GCV choice). The GCV criterion
leads always to a lower regularization than the L-curve choice and then
tends to reduce the smoothing of the solution.
In most of our tests, as in Figs. \ref{fig:good}a, c, d, the GCV 
choice corresponds to a point 
where the errors deduced from the fit become small whereas the L-curve
 criterion
gives a point beyond which  a rapid variation of the fitted parameters with
increasing regularization occurs. The fact that the values of the fitted 
parameters are nearly constant between these two points shows that, for this
level of noise, the method is robust in that sense that the choice of 
the precise value of the regularization parameter is not a crucial point:
any choice that tends
to establish a balance between the propagation of input errors and the 
regularization is able to produce good results.

Let us now look at the behavior of this method for a more
realistic example. For this we take a level of noise similar to
the one given by observers ($k_\sigma=1$) and we build frequency
splittings of sectoral modes by taking into account a
latitudinal variation of the rotation rate in the convection zone close to
 that derived by 2D inversions.
We have set $A=55$ nHz and $B=75$ nHz which are mean values derived from 
observations of the plasma motion at the solar surface (Snodgrass \& Ulrich
\cite{Snod:Ul}). This choice for the
 input rotation law and errors is referred as the
`realistic case' in the following. The Eq.~(\ref{eq:int2D}) with $m=l$
has been used to compute the frequency splittings of sectoral modes and
1D Tikhonov inversions have been performed again in order to infer the 
equatorial rotation rate from Eq.~(\ref{eq:int}).
\begin{figure} \LabelFig{a-d}{fig:bad}
 \psscalefirst
 \centerline{
 \psfig{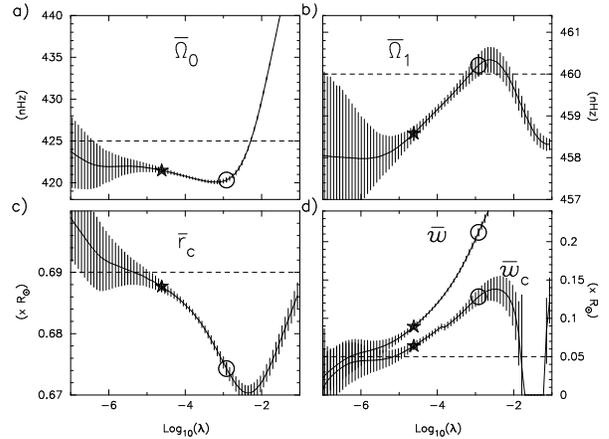}}
 \caption[]{The same as in Fig \ref{fig:good} but with more realistic input 
errors ($k_\sigma=1$) and an input rotation profile with latitudinal variation
in the convection zone ($A=55$ nHz, $B=75$ nHz).}
\end{figure}

Figure \ref{fig:bad} represents the results of these inversions 
in the same form as Fig. \ref{fig:good} and for the same initial 
$erf$-parameters.
There are two essential points to be seen on this figure.
The parameter $\Omega_0$ in Fig. \ref{fig:bad}a is systematically
 under-estimated of about $4$ nHz. 
A detailed analysis shows that this effect is strongly related to 
the introduction of a latitudinal variation of the rotation rate in the 
convection zone. The assumption, used in the 1D inversions, that sectoral 
modes are sensitive only to the equatorial component of the rotation 
rate is not valid for low degree $l$ modes (e.g. Antia et al. \cite{Antia})
, and these modes sound
the deep interior. This may explain some perturbation for the determination
 of the parameter $\Omega_0$ that represents the mean value of the
rotation rate in the radiative interior.
The difference
between splittings of sectoral modes computed from Eq.~(\ref{eq:int}) and 
Eq.~(\ref{eq:int2D}) is below 1 nHz for the observed modes having
their turning points above $0.4R_\odot$.
The  large resulting difference in $\Omega_0$ is due to the 
fact that high $l$ sectoral modes see only  the equatorial rotation
rate and then fix the inferred value $\bar\Omega_1$ equal (or nearly equal as
in Fig. \ref{fig:bad}b)
to the initial value $\Omega_1$ while lower degrees sectoral modes  
are sensitive to the differential rotation of the convection zone 
and this effect can only be accounted for in the inverse rotation law by a 
substantial lowering in $\bar\Omega_0$.
Furthermore we have  checked that two rotation laws with the same $\Omega_1$
but with a difference of $4$ nHz in $\Omega_0$ and 
two rotation laws with the same $\Omega_0$ but with or without latitudinal 
variation in the convection zone, induce a difference of the same order
  in the sectoral modes frequency splittings.

The second important point is that, in Figs. \ref{fig:bad}c, d,
 the estimation $\bar{w}$ of the width of 
the tachocline increases rapidly between the GCV and the L-curve points whereas
its location $\bar r_c$ decreases rapidly from 
$0.688R_\odot$ down to $0.674R_\odot$
As in Fig. \ref{fig:good}d,
the deconvolution made by using averaging kernels tends to correct
 this behavior for the estimation of the width but, in this case,
the GCV choice remains over-estimated for about $0.015R_\odot$
 and the L-curve choice is still very distant from the initial value.
Tests made with different input parameters show that, as in
 Figs. \ref{fig:bad}c, d and  for that level of noise, 
the GCV choice is always better than the L-curve choice
 for the estimation of the location and the width of the tachocline.
This point will be illustrated and discussed in the next section for
the estimation of widths between $0.03$ and $0.11R_\odot$.

\begin{figure} \LabelFig{a-d}{fig:goodk}
\psscalefirst
\centerline{
\psfig{figure=6651f3.ps,height=8cm,angle=-90}}
 \caption[]{The same as in Fig. \ref{fig:good} (`ideal case') 
but for MTSVD (full line) and PP-TSVD (dashed line) methods and against 
the truncation parameter $k$. The L-curve choice for MTSVD method is outside
the plot on panel b.}
\end{figure}
\begin{figure} \LabelFig{a-d}{fig:badk}
\psscalefirst
\centerline{
\psfig{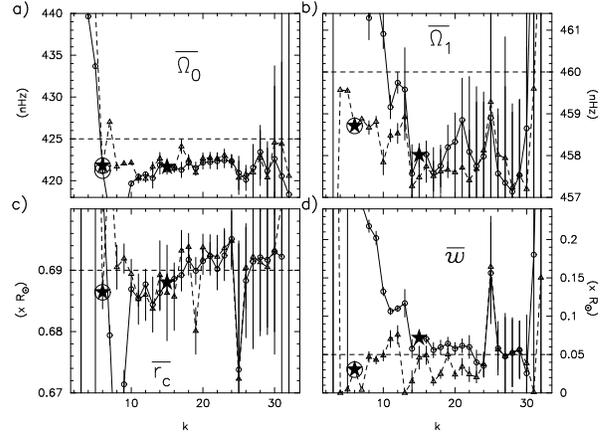}}
 \caption[]{The same as in Fig. \ref{fig:bad} (`realistic case') 
but for MTSVD (full line) and PP-TSVD (dashed line) 
methods and against the truncation parameter $k$. The L-curve choice for
MTSVD method is outside the plot on panels b,c and d.}
\end{figure}
Similar figures (Figs. \ref{fig:goodk}, \ref{fig:badk}) can be plotted 
for MTSVD and PP-TSVD methods where 
the continuous regularization parameter is replaced by the discrete
truncation parameter. Results obtained in the `realistic case' 
(Fig. \ref{fig:badk}) have again a larger dispersion and exhibit
the same systematic deviation 
for the determination of $\Omega_0$. Another interesting point is that, as
shown on Figs. \ref{fig:goodk}d, \ref{fig:badk}d and also in the next section,
 the PP-TSVD
method tends to give  an under-estimation of the width whereas the
MTSVD method tends to give an over-estimation of this parameter. This may
be very useful in order to give a bounded estimation of the true width.
For these two methods, the choice of the 
optimal truncation parameter $k$ through the L-curve criterion needs
the evaluation of the curvature of discrete L-curve. This can be done
carefully by an appropriate 2D curve fitting. Nevertheless our experience
shows that it is difficult to do this systematically with the same fit 
procedure for any level of noise and input rotation law. Furthermore,
when this is done carefully, this choice leads to results for the tachocline 
profile
that are always worse than the ones obtained from the GCV choice. 
Thus, in the following,  results are shown 
only with the GCV criterion for MTSVD and PP-TSVD methods. 

\begin{figure} \LabelFig{a-c}{fig:sol}
\psscalefirst
 \centerline{
 \psfig{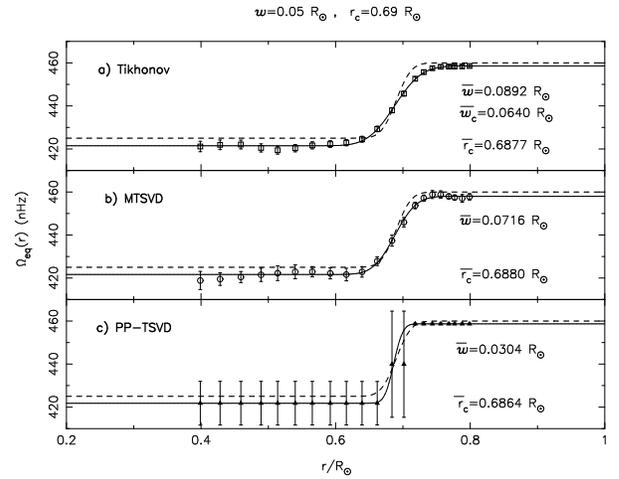}}
 \caption[]{Solutions obtained between $0.4$ and $0.8R_\odot$ 
from the three inverse methods with the GCV
choice of regularization parameters. The input rotation law was the same as in 
Figs. \ref{fig:bad}, \ref{fig:badk} (`realistic case').
 The equatorial component
of the initial law is shown by dashed line whereas the fits of the inverse 
solutions are shown by full lines.}
\end{figure}
Figure \ref{fig:sol} shows the solutions obtained from the three methods
with the GCV choices indicated on Figs. \ref{fig:bad} and  \ref{fig:badk}.
The error bars on the PP-TSVD method (Fig. \ref{fig:sol}c) were obtained
by assuming that the method is linear i.e. the dependence of $\vec{H}$
 (defined in  Eq.~(\ref{eq:non_lineaire})) relatively to the data vector 
$\vec W$ is 
neglected. This is indeed not the case and a Monte-Carlo
approach for estimating errors may be more realistic. We note however that
the two other methods (Tikhonov and MTSVD) are linear only for a given 
regularization parameter. Since  this parameter is chosen through automatic
strategies, it  depends also on the data. Thus, strictly speaking,
these methods are also non-linear methods. Nevertheless, the automatic
choices are built so that they are not too much sensitive to little change
in the data and that justify the linear approximation. 
\begin{figure} \LabelFig{a and b}{fig:avk}
\psscalefirst
 \centerline{
 \psfig{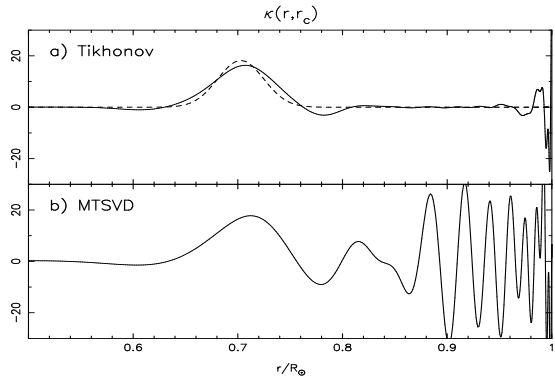}}
 \caption[]{Averaging kernels computed at $r_0=\bar r_c$. For Tikhonov 
method the  dashed line represents the Gaussian approximation of the kernel
used for the local deconvolution of the solution shown on Fig. \ref{fig:sol}a.}
\end{figure}
The corresponding 
averaging kernels computed at $r=\bar r_c$ (Fig. \ref{fig:avk})
 show that whereas the Gaussian 
approximation is rather good for the Tikhonov method, 
the large oscillations in the convection zone obtained for
the MTSVD method make difficult the use of a local deconvolution in that case.

\subsection{Tests for width between $0.03$ and $0.11$ $R_\odot$} 

An important point is to test the ability of a method to give
a good estimation of the $erf$-parameters for a large domain
of variation of the width of the tachocline. We first study in Fig.
\ref{fig:comp} the
behavior of the different methods and automatic strategies between the 
`ideal case'  and the `realistic case' for one realization of input errors.
Then, in Fig. \ref{fig:histo500},
 we have carried out a Monte-Carlo approach in order to have a better 
estimation of the errors on the widths deduced from the fit of the solutions
for the `realistic case'.

\begin{figure} \LabelFig{a-c}{fig:comp}
\psscalefirst
 \centerline{
 \psfig{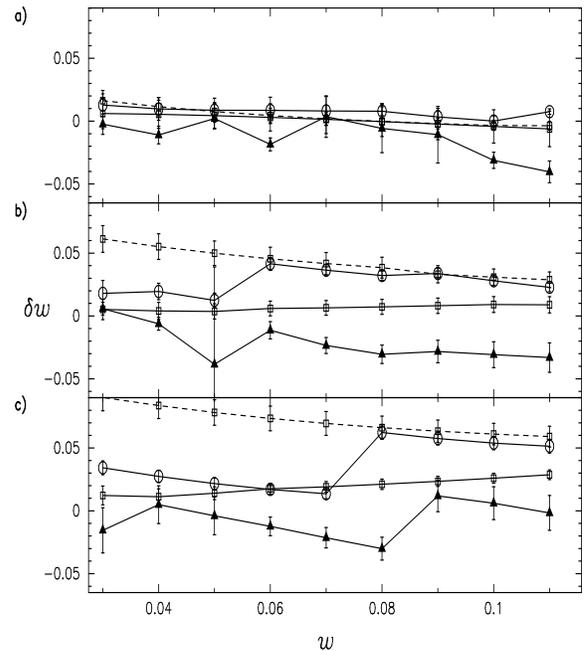}}
 \caption[]{Difference between the inferred width and the initial width 
($\delta w=\bar{w}-w$) against the initial width for PP-TSVD 
(triangles) and MTSVD (circles) methods,  both computed 
with the GCV choice for the truncation parameter.
 Squares are for the Tikhonov method with
GCV criterion (full line) and L-curve criterion (dashed line). For this latter
method we plot the difference between the corrected inferred width and the 
initial width ($\delta w=\bar{w}_c-w$). 
{\bf a} $k_\sigma=10$, $A=B=0$ as in Figs. \ref{fig:good}, \ref{fig:goodk} (`ideal case');
{\bf b} $k_\sigma=1$, $A=B=0$;
{\bf c} $k_\sigma=1$, $A=55$, $B=75$ as in Figs. \ref{fig:bad}, \ref{fig:badk}
(`realistic case')}
\end{figure}

Figure \ref{fig:comp} shows the inferred
width $\bar{w}$ (for MTSVD and PP-TSVD methods) and the corrected 
inferred width $\bar{w}_c$ (for the Tikhonov method)  as  functions
of the initial width $w$ and for one realization of the input errors.
 Figure \ref{fig:comp}a represents the same 
example as Figs.~\ref{fig:good}, \ref{fig:goodk} 
(`ideal case' ), in Fig. \ref{fig:comp}b we increase the 
level of noise ($k_\sigma=1$), and finally we set an input rotation law with a latitudinal
dependence in the convection zone so that
the  Fig. \ref{fig:comp}c is for the same example as Figs. \ref{fig:bad},
\ref{fig:badk} (`realistic case').

In Fig. \ref{fig:comp}a , the results for  
$\bar{w}$ fit the real value within 
$0.02R_\odot$
except for PP-TSVD and widths above $0.9R_\odot$, and the two regularization
procedures (L-curve and GCV) give almost the same result.

The comparison of Figs.~\ref{fig:comp}a and \ref{fig:comp}b clearly indicates
 that the results obtained for  Tikhonov method with the 
L-curve criterion (dashed curves) are very sensitive to the level of noise
and are not adapted to the actual errors of observed data.
The deconvolution method using Tikhonov inversion with GCV
criterion  appears to be the less
 sensitive to the noise level and the most stable for widths between $0.03$
and $0.11R_\odot$. We see again that the results obtained from MTSVD
and PP-TSVD lead respectively to an over-estimation and an under-estimation 
of the real width.
 Figure~\ref{fig:comp}c illustrates the effect of a 
latitudinal dependence of the rotation in the convection zone:
an increasing over-estimation  of  $w$ from the Tikhonov
method with GCV criterion  
and a general larger dispersion of the results.

\begin{figure} \LabelFig{}{fig:histo500}
\psscalefirst
 \centerline{
 \psfig{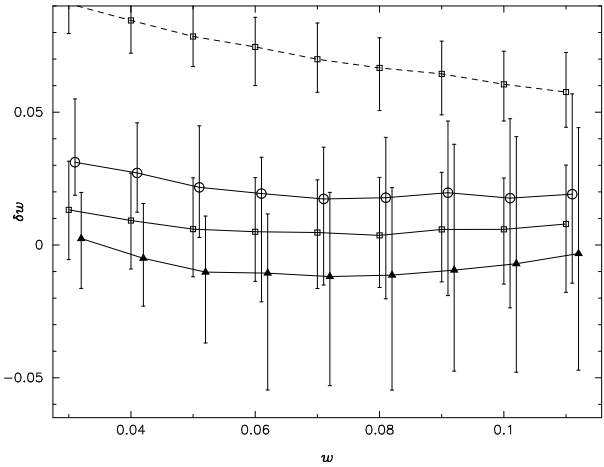}}
 \caption[]{The same as in Fig. \ref{fig:comp}c ('realistic case') 
but each points
is the mean value of the results obtained for 500 realizations of
input errors. Error bars represent a $68.3\%$ confidence interval on $w$.}
\end{figure}

In Fig.~\ref{fig:histo500}, 
we have performed $500$ realizations of input errors
for each initial width and each point shown in this figure  represents
the mean value of the $500$ inferred or corrected inferred widths for a
 given initial width and a given method.
Error bars represent a $68.3\%$ confidence interval which contains the nearest $341$ 
inferred widths from the mean value but they are not necessarily symmetric around this value.
This study
shows that the Tikhonov and PP-TSVD methods with the GCV criterion are
the most reliable for estimating the width in the most realistic case. They 
lead, respectively to an over-estimation and under-estimation
of the width of about $0.01R_\odot$ at the maximum 
for initial widths between $0.03R_\odot$
and $0.11R_\odot$. In that range, the standard deviation obtained for $500$
realizations of input errors is around $0.02R_\odot$ for Tikhonov method
and much larger (up to $0.05R_\odot$ for $w=0.11R_\odot$) for PP-TSVD
method which then appears to be well adapted only to infer very sharp 
transitions. 
Let $\omega_i$ represent 
the widths deduced from $N_r$ hypothetical (non-observed)
realizations of the unknown true width $\hat\omega$.
In the Monte-Carlo method we suppose that we can approximate
the distribution of $(\hat\omega-\omega_i,\ i=1,..N_r)$ 
by the distribution of $(\omega_o-\tilde\omega_i,\ i=1,N_r)$ 
where $\omega_o$ is the width deduced from the observed dataset 
and $\tilde\omega_i$ are the 
widths deduced from datasets built by setting $\hat\omega=\omega_o$ 
in the model. As we can not insure that $\omega_o$ is very 
close to $\hat\omega$,
the underlying assumption is that, in the range of uncertainty concerning
$\hat\omega$ (say $0.03-0.11R_\odot$),
 the way in which errors propagate through
the inverse process does not vary rapidly 
(see e.g. Press et al. \cite{NumRec}).
The fact that, in Fig. \ref{fig:histo500}, error bars
 grow rapidly with the initial width for PP-TSVD method makes difficult
the use of the Monte-Carlo results for estimating the statistical behavior
of this  method.    
There are nevertheless two factors  
that may introduce bias in these estimations 
of  the errors on the inferred widths. First, the existing 
correlations between the inferred rotation values obtained at two different 
radius are not taken into account in the  fit of the solution by an 
$erf$-function. 
Secondly, for the PP-TSVD method, the non-linearity of the method is not 
taken into account in the estimation of the propagation of noise through 
the inverse process. Making the fit in the right way, i.e. taking into account
correlations, may lead to a lower dispersion of the results and then  
our estimation of the error on the inferred widths may be  over-estimated. 
Nevertheless, the effects of these two approximations  are not easy to
estimate a priori and  need a more complete analysis in future work.

\begin{table*}
\caption[]{Inferred $erf$-parameters obtained from LOWL data. The L-curve 
criterion has not been used for methods with discrete truncation parameters.}
\begin{flushleft}
\begin{tabular}{llllllll}
\hline\noalign{\smallskip}
Methods & \multicolumn{2}{l}{$\bar\Omega_0$ (nHz)} & &\multicolumn{2}{l}{$\bar\Omega_1$ (nHz)}& 
$\bar r_c/R_\odot$ & $\bar w_{(c)}/R_\odot$ \\
\noalign{\smallskip}
\cline{2-3}\cline{5-6}
\noalign{\smallskip}
        & GCV & L-curve                            & & GCV & L-curve                          &
GCV                   & GCV                         \\ 
\noalign{\smallskip}
\hline
\noalign{\smallskip}
Tikhonov & $429.3\pm 0.5$ & $427.9\pm 0.3$ & &$457.7\pm 0.3$ & $460.4\pm 0.4$ & $0.693\pm 0.002$ & $0.067\pm 0.010$ \\
MTSVD	 & $429.4\pm 0.7$ & - & &$457.0\pm 0.5$ & - &$0.693\pm 0.003$ & $0.062\pm 0.009$ \\
PP-TSVD  & $429.6\pm 0.2$ & - & &$456.4\pm 0.3$ & - &$0.693\pm 0.009$ & $0.031\pm 0.017$ \\
\noalign{\smallskip}  
\hline
\end{tabular}
\end{flushleft}
\end{table*}

\section{Results for LOWL data}\label{sec:lowl}

\begin{figure} \LabelFig{}{fig:sol_lowl}
\psscalefirst
\centerline{
\psfig{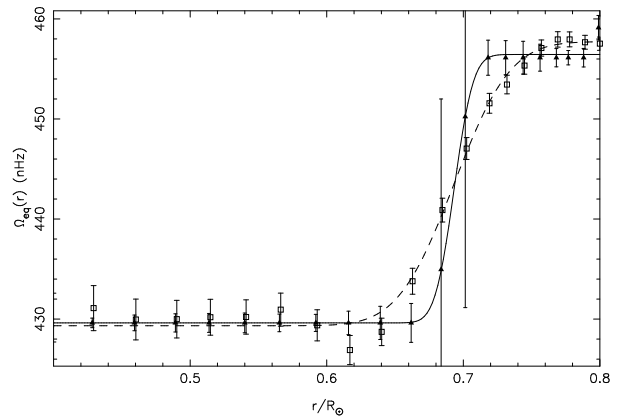}}
 \caption[]{Equatorial tachocline profiles  obtained from LOWL data by 
PP-TSVD (triangles) and Tikhonov (squares) methods with GCV criterion.
 Error bars represent the 
$1\sigma$ errors estimated on the solution by assuming the linearity
of the inversions. The full and dashed curves represent respectively the
fit of the PP-TSVD and Tikhonov solutions by an $erf$-function between
$0.4$ and $0.8R_\odot$.}
\end{figure} 
This section gives the results obtained from the two years (2/26/94-2/25/96) 
observations by  the LOWL instrument in Hawaii (Tomczyk
et al. \cite{Tomczyk}; Corbard et al. \cite{meAA}). These data contain $1102$
modes with degrees up to $l=99$ and frequencies between $1200$ and $3500$ 
$\mu$Hz. For each mode $(n,l)$, individual splittings are given by, 
at best, five 
a-coefficients of their expansion on orthogonal polynomials defined by 
Schou et al. (\cite{SCDT}). For this work, we assume that the previous 
simulations provide an estimation of the bias introduced by the methods 
and we use these values in order to correct the inferred tachocline 
parameters.  This supposes
the closeness of the model used in the simulation to the reality and a good
estimation of the errors in the data.
Furthermore,
we use the  sum of odd 
a-coefficients as a first approximation for the sectoral splittings i.e. 
$\Delta\nu_{nll}\simeq a^{nl}_1+a^{nl}_3+a^{nl}_5$.
This approximation is exact for all the rotation laws such that 
$a^{nl}_{2j+1}=0\ \forall\ j>2$ (which is the case for the rotation
laws Eq. \ref{eqn:law} used in our model). When this is not the case
the latitudinal kernel associated to $a^{nl}_1+a^{nl}_3+a^{nl}_5$
is less peaked at the equator than the one associated to
the sectoral splittings (i.e. $P_l^l(\cos\theta)^2\sin\theta$,
see Sect.~\ref{sec:Model}) and thus $\hat\Omega_1$ represents
a latitudinal average of the rotation in a larger domain 
around the equator. However the kernel associated to the sum of three
a-coefficients is less $l$-dependent.

Results obtained by the three methods are summarized in Table 1. 
They are in very good agreement for the location of the tachocline and
the mean values of the rotation rate in the radiative interior and convection 
zone but more dispersive concerning the determination of the width. The tests
discussed above have shown that this may be related to the level of noise
contained in the data.  
The  equatorial tachocline profiles 
obtained by Tikhonov and PP-TSVD methods with GCV criterion are shown
in Fig. \ref{fig:sol_lowl}. 
According to the previous sections, we  use the GCV choice
 in order to
infer the location and the width of the equatorial tachocline. Nevertheless, 
for $\Omega_0$ and $\Omega_1$ the L-curve 
choices may be useful in order to see the amplitude of the variation of
 the inferred
parameters against the regularization parameter.
The errors cited in this Table  are just the result of the fit of the solution
by the $erf$-function. The variation of the inferred $erf$ parameters
against the regularization,
as shown by Fig. \ref{fig:lowlreg} for the Tikhonov method, and the previous
Monte-Carlo simulations 
can help us to estimate error bars that may be more realistic.

\begin{figure} \LabelFig{}{fig:lowlreg}
\psscalefirst
\centerline{
\psfig{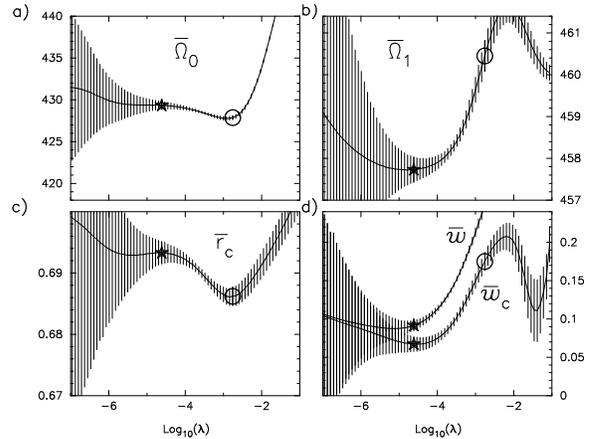}}
 \caption[]{ Variation of the inferred parameters  $\bar\Omega_0,  
\bar\Omega_1, \bar r_c,\bar{w}$ and $\bar{w}_c$ as a 
function of the logarithm of the regularization parameter 
for the Tikhonov inversion of LOWL data. Graph markers have the same 
meaning as in Fig. \ref{fig:good}. The L-curve choice of $\bar{w}$ 
is outside the plot on panel d.}
\end{figure}

 Figure \ref{fig:lowlreg}a shows that 
the evaluation of the mean value of the rotation
rate in the radiative interior ($\hat\Omega_0$) 
is not much sensitive to the regularization.
Nevertheless, we have shown in Sect.~\ref{sec:key} that this 
parameter  tends to be 
systematically under-estimated
of about $4$ nHz because of the influence of the latitudinal variation
of the rotation in the convection zone on the low $l$ sectoral splittings.
For the sum $a_1^{nl}+a_3^{nl}+a_5^{nl}$ the latitudinal kernel
is less $l$-dependent so that this systematic offset may be smaller than
$4$ nHz. We take this effect into account by increasing the estimation of the 
error and  
our final interval for this parameter becomes:
 $427.5\le\hat\Omega_0\le434.5$ nHz.
The mean value of the equatorial rotation rate in the convection zone
is less subject to systematic errors but may be under-estimated by the
GCV choice (c.f. Figs. \ref{fig:bad}b, \ref{fig:badk}b, \ref{fig:sol}). 
The difference between the
GCV choice and the L-curve choice is about $3$ nHz on Fig. \ref{fig:lowlreg}. 
Thus we estimate
that $\hat\Omega_1=459.0\pm 1.5$ nHz.        
We note that 
we do not attempt to use the points of the solution found under $0.4R_\odot$
or above $0.8R_\odot$ (c.f. Fig. \ref{fig:sol_lowl}). 
Therefore $\hat \Omega_1$ does not take  into account
the eventual rapid variation of the rotation near the surface or at 
$0.9R_\odot$ (Antia et al. \cite{Antia}) and $\hat \Omega_0$ is not 
sensitive to the core rotation.
The ratio $q=\hat\Omega_0/\hat\Omega_1$ obtained from helioseismic data 
is an important test for the theories of the tachocline dynamics.
Spiegel and Zahn's (\cite{tacho1}) theory leads to $q=0.90$ whereas
Gough's (\cite{Gough85}) one leads to $q=0.96$. Our results give
$0.93<q<0.95$ which is intermediate between the two theoretical estimates.
Similar results have already been pointed out
by Gough \& Sekii (\cite{tacho2}).

For the estimation of $\hat r_c$, we find in Fig. \ref{fig:lowlreg} that the 
L-curve criterion leads to a lower value than the GCV criterion as we
had found in Fig. \ref{fig:bad}. As discussed in Sect.~\ref{sec:key}, 
we think that the GCV choice is more reliable
but may lead to an under-estimation of about $0.002R_\odot$.
Therefore our final estimation for the location of the center 
of the tachocline in the equatorial plane is: 
$\hat r_c=0.695\pm 0.005R_\odot$. 
This value, estimated in the equatorial plane, is intermediate between 
the two values previously obtained
by forward methods (c.f. Table 2). We note however that whereas our work just
look for the equatorial component of the tachocline,  the previous 
works assume that the solar tachocline presents the same profile at
any latitude. This may lead to bias if, as suggested by Charbonneau et al.
(\cite{char}) from LOWL data, the tachocline is prolate i.e. is located 
deeper at the equator than at higher latitudes.

The tests discussed in the previous sections show that the L-curve choice
is not reliable for the estimation of the width and suggest three ways for
estimating the width of the tachocline from GCV criterion:

First, the true value is supposed to lie between the MTSVD and PP-TSVD
estimations. That gives $0.031R_\odot\leq\hat{w}\leq 0.062R_\odot$.

Secondly, for the Tikhonov method, since the error bars have 
roughly of the same amplitude in the whole range $0.03-0.11R_\odot$ of
initial widths (Fig. \ref{fig:histo500}) , 
we can use the Monte-Carlo simulation. Near $w=0.07R_\odot$
 (the inferred value reported in Table 1 being $\bar{w}=0.067$),
Fig. \ref{fig:histo500} shows that    
the Tikhonov method leads in mean 
to a systematic over-estimation of about $0.005R_\odot$
with a $68.3\%$ confidence interval around $\pm 0.02R_\odot$. 
Thus we obtain by this way 
$\hat{w}\simeq 0.062\pm 0.020R_\odot$.

Thirdly, the PP-TSVD method is though to produce, in mean, an under-estimation
of the width of about $0.01R_\odot$ but with a larger dispersion of 
the results for the large widths so that we are  not allowed
 to use straightforwardly
our Monte-Carlo simulation. The $68.3\%$ confidence intervals plotted in
Fig. \ref{fig:histo500} indicate that the
PP-TSVD method can lead to an inferred width around $0.03R_\odot$ (which 
is the value obtained from LOWL data) for  initial
widths up to $0.08R_\odot$. Therefore the interpretation of the result
obtained by this method is not easy. This may indicate that the method is
 better suited to the search of transition zones known a priori to be very 
thin (searching for a width lower than $0.05R_\odot$ for example).
Nevertheless, all the above discussions indicate
 $0.020\le\hat{w}\le 0.070R_\odot$ as  a reasonable 
interval for the true width, deduced from
PP-TSVD method.

All these approaches are globally consistent but lead to a relatively large
dispersion of the results. Therefore our final 
estimation of the width
of the solar tachocline in the equatorial plane is: 
$\hat w=0.05\pm0.03R_\odot$.
This estimation is  in very good agreement with the result obtained by
Charbonneau et al. (\cite{char}) 
and remains compatible with the value given by
 Kosovichev (\cite{koso}) (c.f. Table 2).

\begin{table}
\caption[]{Comparison of our results with previous forward analysis.
Charbonneau et al. (\cite{char}) and our work
are for the same  LOWL dataset (2/26/94-2/25/96) whereas Kosovichev (\cite{koso})
has used the 1986-90 BBSO datasets.}
\begin{flushleft}
\begin{tabular}{lll}
\hline\noalign{\smallskip}
& $\hat{r_c}/R_\odot$ & $\hat{w}/R_\odot$ \\
\noalign{\smallskip}
\hline\noalign{\smallskip}
This work          & $0.695\pm 0.005$ & $0.05\ \pm 0.03 $\\ 
Charbonneau et al. & $0.704\pm 0.003$ & $0.050\pm 0.012$\\
Kosovichev         & $0.692\pm 0.005$ & $0.09\ \pm 0.04 $\\
\noalign{\smallskip}  
\hline
\end{tabular}
\end{flushleft}
\end{table} 

\section{Conclusions}

This work presents an analysis of the determination of the characteristics
of the tachocline at the equator by three different inverse methods. They are
 applied to the inversion
 of the splittings of the sectoral modes estimated as the sum of the three
 first odd coefficients of the expansion of  the splittings in orthogonal polynomials defined by Schou et al. (\cite{SCDT}).  Two 
different choices of regularization parameters, the GCV and L-curve criteria,
have been compared. Tests with artificial
 rotation laws have shown that in all cases the GCV criterion is less 
sensitive to the error level than the L-curve one  and gives
 better results with low bias and dispersions in the range $0.03-0.011R_\odot$
of searched widths. This choice of the GCV criterion is in agreement with
 Barett (\cite{Barrett}) and Thompson(\cite{ThompsonAM1}) in another context. 
Hansen (\cite{L-curve}) has shown that the GCV criterion is less adapted
to highly correlated errors than the L-curve one. Our work may indicate in 
turn that we can neglect, as it has been done, the unknown correlation in 
LOWL data.  

 Concerning the thickness of the tachocline, it appears  that the MTSVD
 and PP-TSVD inversions give respectively an upper and lower estimate while
  the
 Tikhonov method corrected by deconvolution gives the most reliable
 determination. 
We have estimated the systematic effect of the latitudinal dependence of
 the rotation
in the convection zone on the determination of the thickness of the
 tachocline and the rotation in the radiative interior.
 We have  shown how the performance of the methods will
 be improved by  lowering the level of noise in the data.

The methods have been applied to the LOWL two years dataset leading
to an estimation of the position $\hat r_c=0.695\pm 0.005R_\odot$
and the thickness  $\hat w=0.05\pm 0.03R_\odot$ of the equatorial tachocline.
In addition,
we have obtained an estimation of the equatorial rotation 
$\hat \Omega_0$ below the 
convection zone and above $0.4 R_\odot$ such that: 
$427.5\le\hat \Omega_0\le 434.5$ nHz and $\hat \Omega_1$
from the top of the convection zone up to $0.8 R_\odot$ such that
$\hat \Omega_1=459.0\pm 1.5$ nHz. Assuming 
that the rotation in the radiative interior is independent
of latitude,
this leads to a ratio $\hat \Omega_0/\hat \Omega_1$
 between  $0.93$ and $0.95$ which is intermediate between the two theoretical 
predictions.
 
Our results  
 for the location and thickness of the equatorial tachocline 
 are in agreement with the forward analysis of 
 Charbonneau et al. (\cite{char}) and with those
 of Basu applied on  BBSO and GONG datasets (Basu \cite{Basu}) 
using a different parameterization of the tachocline.
The forward analysis can be viewed as non-linear least-squares
 methods
(least-squares methods  because of the use of the $\chi^2$ criterion and 
non linear because of the models used for the rotation profile)
but using only a very few number of parameters
 (Charbonneau et al. (\cite{char}) use six parameters, Basu (\cite{Basu}) 
three and
Kosovichev (\cite{koso}) only two). This kind of methods 
 depend thus strongly on our knowledge of the global rotation
profile which can be reached only by inversion techniques. In particular, 
in the above-cited works  the latitudinal dependence of the rotation is fixed 
(as in 1.5D inversions). In this work, we have tried to investigate the 
amount of
informations about the tachocline that we can extract directly from the global 
inversions without a-priori knowledge (except for the regularization) 
on the rotation profile. There are less assumptions in this approach, 
and thus the tachocline parameters may be less constrained.  
The fact that the two approaches lead to similar results 
indicates in turn that the hypothesis used in the forward analysis are
probably not too strong and are
well adapted to the problem of inferring the tachocline from actual data.

One of the  interest of this work was our first attempt to use an
inverse method with non-linear regularization in helioseismic case.
The  PP-TSVD method leads to a very large dispersion of 
the results for widths above $0.05R_\odot$ and then is difficult to interpret
with actual data. Some efforts, in future work,  
should be useful to improve this kind of methods and the 
interpretation of their results taking into account their non-linearity.

\begin{acknowledgements}

We gratefully acknowledge T. Sekii for illuminating discussions and comments,
S. Tomczyk and J. Schou for providing the LOWL data and the anonymous referee
for constructive comments.

\end{acknowledgements}

\appendix
\sectionApp{Details of the three inverse methods used}\label{app:methods}

We discretize Eq.~(\ref{eq:int}) by: 

\begin{equation}
\vec W=\vec R\vec\Omega
\end{equation}
where we have defined:

\begin{equation}
\vec W\equiv(W_i)_{i=1,N} \\
W_i=\Delta\nu_{nll}+\delta\nu_{nll},\ i\equiv(n,l),
\end{equation}
$N$ being the number of modes $(n,l)$ ($N=1102$ for LOWL data)
and $\delta\nu_{nll}$ a normally 
distributed noise with a standard deviation defined in Eq.~(\ref{eq:err}).
We search the solution $\bar\Omega(r)$ as a piecewise linear function
of the radius by setting:
\begin{equation}\label{eq:sol}
\bar\Omega(r)=\sum_{p=1}^{N_p} \omega_p\varphi_p(r) \\ 
\vec\Omega\equiv (\omega_p)_{p=1,N_p}
\end{equation}
where $\varphi_p(r)$, $p=1,N_p$ are piecewise straight lines ($N_p=50$ in this
work) such that:

\begin{equation}
\forall p=1..N_p,\ \exists\ r_p\ \in [0.,1.]\ /\ \bar\Omega(r_p)=\omega_p
\end{equation}
where $r_p,\ p=1..N_p$ are fixed break points distributed according to the
density of turning points of modes (Corbard et al., \cite{meAA}).
The matrix $\vec R$ is then defined by: 
\begin{equation}\label{eq:discret}
\vec R\equiv (R_{ip})_{{i=1,N\atop p=1,N_p}} \\
R_{ip}=\int K_{nl}(r)\varphi_p(r) dr
\end{equation}

For all the inverse methods discussed in this work,
 the aim is to find a solution that is able to produce a good
fit of the data in chi-square sense.
 Unfortunately, the solution of this problem
is not unique and  allows
oscillatory solutions that are not physically acceptable. So, we have to
define a quantity that measures
 the smoothness of the solution and to insure that
the final solution is sufficiently smooth to be acceptable.

For any solution $\vec\Omega$, we define the $\chi^2$ value by:

\begin{equation}
\chi^2(\vec\Omega)=\|\vec P(\vec R\vec\Omega-\vec W)\|_2^2
\end{equation}
where $\vec P=diag(1/\sigma_{nl})$ and we define two measures
of the smoothness of the solution $\vec\Omega$ by:

\begin{equation}
\beta_i(\vec\Omega) =\|\vec L \vec \Omega\|_i,\ i=1,2
\end{equation} 
where the vector i-norms $\|.\|_i$ are defined by
$\|\vec{x}\|_i=(\sum_p |x_p|^i)^{1/i}$ and 
$\vec L$ is a discrete approximation of the first derivative operator such
that:

\begin{equation}
\beta_1 \propto \int \left|{\partial\Omega(r)\over\partial r}\right|dr \\
\beta_2^2 \propto \int \left({\partial\Omega(r)\over\partial r}\right)^2 dr
\end{equation}  

\subsection{Tikhonov solution}\label{app:Ti}

The so called Tikhonov solution $\vec\Omega_\lambda$ solves the problem:

\begin{equation}\label{eq:Ti}
\min_{\vec\Omega}\ (\chi^2(\vec\Omega)+\lambda \beta_2^2(\vec\Omega)),
\end{equation}
where $\lambda >0$ is the continuous regularization parameter.
In order to compare this method to the two other ones, it may be
interesting to reformulate the problem as follow:
For any $\lambda$ we can show that there exist a value $\alpha(\lambda)$ 
for which
$\vec\Omega_\lambda$ is the solution of the problem:

\begin{equation}
\min_{\vec\Omega\in S_\lambda} \beta_2(\vec\Omega);\  \\ 
 S_\lambda=\{\vec\Omega\ /\ \|\vec P(\vec R\vec\Omega-\vec W)\|_2
\leq \alpha(\lambda)\}
\end{equation}
The computation of these solutions for different regularization
parameters have been carried out by using a generalized singular
value decomposition of the pair $(\vec R,\vec L)$ as explained
and discussed extensively in Christensen-Dalsgaard et al. (\cite{GSVD}). 

\subsection{MTSVD and PP-TSVD solutions}\label{app:M-PPTSVD}

These methods are based on the SVD of the $N\times N_p$ ($N>N_p$) matrix $\vec R$ 
which can be written:

\begin{equation}\label{eq:svd}
\vec R=\sum_{i=1}^r {\vec u}_i \Lambda_i {\vec v}_i^\top
\end{equation}
where $r\le N_p$ is the rank of $\vec R$.
 The singular vectors are orthonormal, $\vec{u}_i^\top\vec{u}_j=
\vec{v}_i^\top\vec{v}_j=\delta_{ij}$ for $i,j=1,r$, and the singular values
$\Lambda_i$ are such that: $\Lambda_1\geq\Lambda_2\geq ...\geq\Lambda_r> 0$,
$\Lambda_{r+1},..,\Lambda_{N_p}=0.$
We then define the the TSVD of $\vec R$ as the matrix $\vec R_k$ built from
Eq.~(\ref{eq:svd}) but neglecting the $N_p-k$ smallest singular values. 

\begin{equation}\label{eq:tsvd}
\vec R_k=\sum_{i=1}^k {\vec u}_i \Lambda_i {\vec v}_i^\top
\end{equation}
The integer
$k<r$ is called the truncation parameter. It acts
 as a regularization parameter 
by eliminating the oscillatory behavior of the singular vectors 
associated with the smallest singular values.
According to Eq.~(\ref{eq:tsvd}), the rank of the matrix $\vec R_k$ is $k<r$
and then the problem of minimizing the quantity $\|\vec P(\vec R_k\vec\Omega-\vec W)\|_2 $
has not an unique solution and we have to use our smoothness criteria
to select a physically acceptable solution among the set of solutions
defined by:

\begin{equation}
 S_k=\{\vec\Omega\ /\ \|\vec P(\vec R_k\vec\Omega-\vec W)\|_2 
= \mbox{minimum} \}
\end{equation}
With these notations, the so-called MTSVD solution $\vec\Omega_k^m$ is 
defined by:

\begin{equation}
\vec\Omega_k^m=\arg \min_{\vec\Omega\in S_k} \beta_2(\vec\Omega) \\ 
\end{equation}
whereas the so-called PP-TSVD solution $\vec\Omega_k^p$ is defined by:

\begin{equation}
\vec\Omega_k^p=\arg \min_{\vec\Omega\in S_k} \beta_1(\vec\Omega) 
\end{equation} 
The algorithms for computing these solutions are presented in Hansen et
al. (\cite{MTSVD2}) and Hansen \& Mosegaard (\cite{PPTSVD}) respectively.

We just recall some important properties of the PP-TSVD solution:
For any $k<r$ the vector $\vec L\vec\Omega_k^p$ has at the most
$k-1$ non zero elements. As the matrix $\vec L$ is a discrete approximation
of the first derivative, this means that the solution vector $\vec \Omega_k^p$
consists on $k_b\leq k$ constant blocks. From Eq.~(\ref{eq:sol}) it follows
that the inferred rotation $\bar\Omega(r)$ itself is obtained as a 
piecewise constant functions with a maximum of $k$ pieces.
 The $k_b-1$ break points
of this solution are selected by the procedure among the $N_p$ initials
break points $r_p$. Therefore this inversion is able to produce 
a discontinuous solution without fixing a-priori the location of
the discontinuity.
Finally, we note that the solution $\vec\Omega_k^p$ obtained by this non-linear 
method
can always be computed by applying a matrix $\vec H$,
 to the data but
this matrix is also a function of the data i.e. $\vec H=\vec{H}(\vec{W})$.
Thus we have:

\begin{equation}\label{eq:non_lineaire}
 \vec\Omega_k^p=\vec{H}(\vec{W})\vec{W}.
\end{equation}


\end{document}